\documentclass[aj]{emulateapj-rtx4}

\usepackage{amssymb}
\usepackage{soul}
\usepackage{graphicx}
\usepackage{amsmath}
\usepackage[usenames]{color} 
\usepackage{ulem}
\usepackage{lscape} 
\usepackage{rotating}
\newcommand\degrees{^\circ}

\newcommand\ie{{\it i.e.~}}

\shorttitle{}

\begin{document}

\title{Main-belt Asteroids in the K2 Engineering Field of View}

\author{R. Szab\'o\altaffilmark{1}}
\email{rszabo@konkoly.hu}
\author{K. S\'arneczky\altaffilmark{1,2}}
\author{Gy. M. Szab\'o\altaffilmark{1,2,3}}
\author{A. P\'al\altaffilmark{1,4}}
\author{Cs. P. Kiss\altaffilmark{1}}
\author{B. Cs\'ak\altaffilmark{2,3}}
\author{L. Ill\'es\altaffilmark{4}}
\author{G. R\'acz\altaffilmark{4}}
\author{L. L. Kiss\altaffilmark{1,2}}
\altaffiltext{1}{Konkoly Observatory, Research Centre for Astronomy and Earth Sciences, Hungarian
Academy of Sciences, H-1121 Budapest, Konkoly Thege Mikl\'os \'ut 15-17, Hungary}
\altaffiltext{2}{Gothard-Lend\"ulet Research Team, H-9704 Szombathely, Szent Imre herceg \'ut 112, Hungary}
\altaffiltext{3}{ELTE Gothard Astrophysical Observatory, H-9704 Szombathely, Szent Imre herceg \'ut 112, Hungary}
\altaffiltext{4}{E\"otv\"os Lor\'and Tudom\'anyegyetem, H-1117 P\'azm\'any P\'eter s\'et\'any 1/A, Budapest, Hungary}

\begin{abstract}
Unlike NASA's original Kepler Discovery Mission, the renewed K2 Mission will stare at the plane of the Ecliptic, observing each field for approximately 75 days. This will bring new opportunities and challenges, in particular the presence of a large number of main-belt asteroids that will contaminate the photometry. The large pixel size makes K2 data susceptible to the effect of apparent minor planet encounters. Here we investigate the effects of asteroid encounters on photometric precision using a sub-sample of the K2 Engineering data taken in February, 2014. We show examples of asteroid contamination to facilitate their recognition and distinguish these events from other error sources. We conclude that main-belt asteroids will have considerable effects on K2 photometry of a large number of photometric targets during the Mission, that will have to be taken into account. These results will be readily applicable for future space photometric missions applying large-format CCDs, such as TESS and PLATO.     
\end{abstract}


\section{Introduction}

\begin{figure*}
\centering\includegraphics[width=8cm,angle=270]{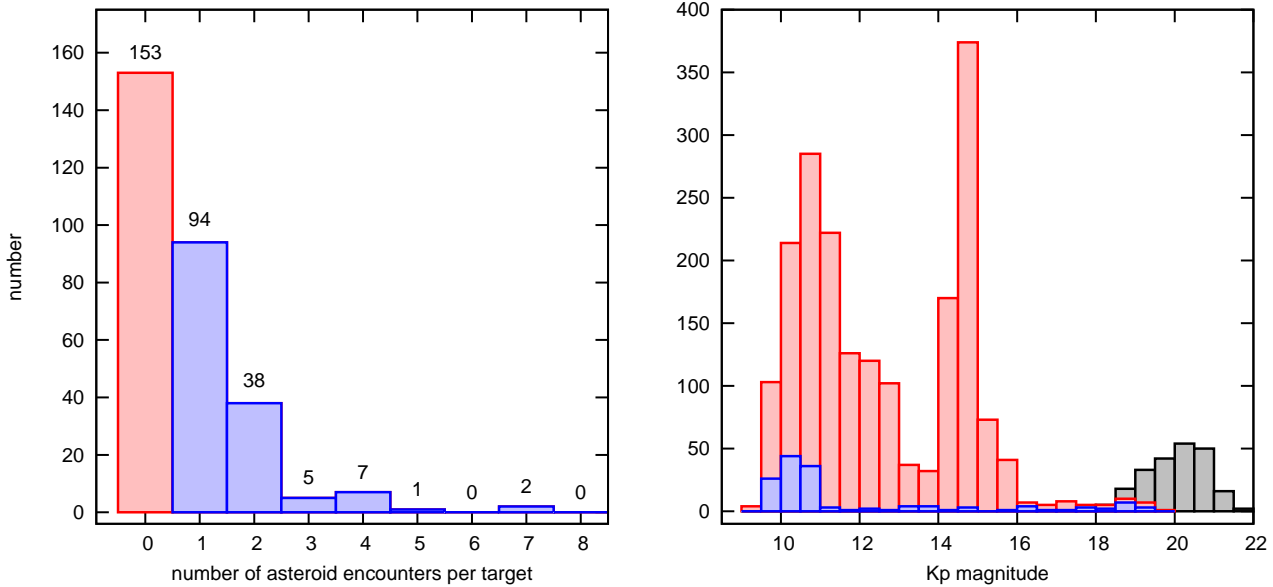}
\caption{{\bf Left:} Number of asteroid encounters in our 300-strong sub-sample. Half of the targets experienced zero encounters (red bar), while many K2-E2 target stars had one or more minor planet approach events (blue bars). {\bf Right:} Magnitude distribution of the K2-E2 targets (red columns). Targets with all magnitudes in the range 9-20 were selected, but some magnitude bins are more strongly populated. Our sample of 300 targets are shown by the blue (smaller) columns. The brightness of the asteroids encountering our targets are faint, shown by the black histogram.}
\label{fig:hist12}
\end{figure*}

The Kepler spacecraft retained much of its operational characteristics \citep{borucki2010} in the K2 Mission, except its pointing (and consequently photometric) accuracy, and the way its pointing is maintained \citep{howell2014}. The space telescope observed a star field in Pisces (center coordinates: $\alpha = 359\degrees$, $\delta$= -2$\degrees$) in February 2014. The primary goal of the K2 Two-Wheel Concept Engineering Test (hereafter K2-E2) was to test the performance of the telescope in fine guidance mode. The observed 2079 long-cadence (30-min sampling), and 17 short-cadence (1-min) targets were made available to the scientific community. In the K2 Mission, 50x50-pixel sub-frames containing the targets are stored, much larger than the usual optimal mask applied in the nominal mission to avoid any flux loss due to the drift of the telescope. These will be optimized as the characteristics of the pointing accuracy and necessary correction will be worked out later during the Mission. An approximately 1-pixel variation in the position of the stars is clearly seen during the 8.9-day long K2-E2 data. In addition, the boresight was shifted by multiple pixels at BJD = 56695.36, ie. 2.3 days after the start of the run.

During the first inspection of the public data, we noticed that a large number of asteroids cross the 50x50 pixel masks, which is obviously due to the selection of the fields close to the Ecliptic plane. Based on the brightness of these objects, the overwhelming majority should be main-belt asteroids. Most of the objects are seen in 6-8 consecutive long-cadence frames as elongated stripes, many of them approaching the main photometric target. A large fraction of them actually cross the PSF of the stars. Since Kepler has large pixel size (3.95$\arcsec$/ pixel) to gather as many photons as possible, the target stars have increasingly higher chance to be blended by asteroids from time to time,  than traditional ground-based telescope-detector systems (typically 0.1-0.5$\arcsec$/pixel for minor planet astrometry observations). This problem will be relevant in future space photometric missions with a similar scope and design to Kepler, like TESS \citep{ricker2014} and PLATO \citep{rauer2014}. In this paper we use the term {\it encounter} when an asteroid enters a user-defined pixel mask of the K2 target, since the probability of an {\it occultation} (when the asteroid passes in front of the stellar disc as seen from Kepler) is extremely low. 

This work is devoted to investigate the following questions: 

- How many asteroid crossings are seen in the original K2-E2 data?

- What is the photometric effect that these crossings might cause and how to take them into account when high-precision photometry is derived?

- What photometric methods are best suited to derive accurate photometry of these Solar System objects?

- Can we identify these main-belt minor planets taking into account Kepler's position in the Solar System? Are there any new among them?

The structure of the paper is the following. In Sec.~\ref{sample} we discuss the selection of our sample and the number of asteroid encounters. Then in Sec.~\ref{effects} the photometric effects of main-belt asteroids on K2-E2 light curves are investigated. In Sec.~\ref{ident} we introduce our methods that led to the identification of these minor bodies (the details of the method will be published in a subsequent paper). In Sec.~\ref{phot} we demonstrate the possibilities of accurate photometry of moving targets (\ie main-belt asteroids) in the K2 Mission. Finally we close the paper with summarizing our results in Sec.~\ref{sum}.

\section{The sample and encounter statistics}\label{sample}

In order to create a statistically meaningful sample, we selected the first 300 LC targets of the available 2079 long cadence targets. We created short animations from the fits frames and these 'movies' were scrutinized for asteroid contamination. Each movie was inspected visually by two persons  independently. By visual inspection we counted those encounter events, where the asteroid 'crossed' or 'touched' a pixel that belonged to the target star. Later, {\it a posteriori}, we established that before and after the encounter event these pixels had at least a flux value $3\thinspace \sigma$ above the background level.

The minor bodies were identified by the method outlined in Sec.~\ref{ident}.  
That method enabled us to check those asteroids that we missed by the visual inspection. We found that we recovered all the encounter events that belonged to asteroids brighter than 21 magnitude. Only a few, fainter asteroid were missed in the 21 -- 21.5 magnitude range.

During the nine days of engineering observations we found that 147 stars suffered encounter(s) out of our sample of 300, which amounts to 49\%, although the number of stars suffering significant effect is much less. 
However, the total number of asteroid encounter events is much higher in our sub-sample of 300 stars, namely 232. Thus, multiple encounters were frequent during 9-day observational period, as shown in a histogram in the left hand panel of Fig~\ref{fig:hist12}. As we show with the first red bar, about half of the stars were  not approached by visible minor planets during the observations, 94 stars had one asteroid, 38 targets had two, and multiple approaches also happened: two stars had as many as seven asteroid fly-bys.

The magnitude distribution of all the K2-E2 long cadence targets is shown in the right hand panel of Fig~\ref{fig:hist12}, again in a form of a histogram (red columns). Smaller, blue bars denote our sub-sample that was scrutinized for asteroid encounters. Both the total sample and our sub-sample covers the whole apparent brightness range (in Kp, the Kepler magnitude system) of 9-20, with a strong increase between Kp 14-15 in the original sample, which reflects {\it a priori} target selection criteria instead of the underlying magnitude distribution of the population of stars in the K2-E2 field. There were only a few relatively bright asteroids in the selected pixel masks, we counted 15 in the 15.5-18.5 magnitude range. All the others are fainter with a clear peak around 20-21 apparent ($V$) magnitudes. The magnitude distribution of the asteroids is shown in the right hand panel of Fig~\ref{fig:hist12} by the black columns. 

\begin{figure}[t]
\centering\includegraphics[width=7.2cm,angle=270]{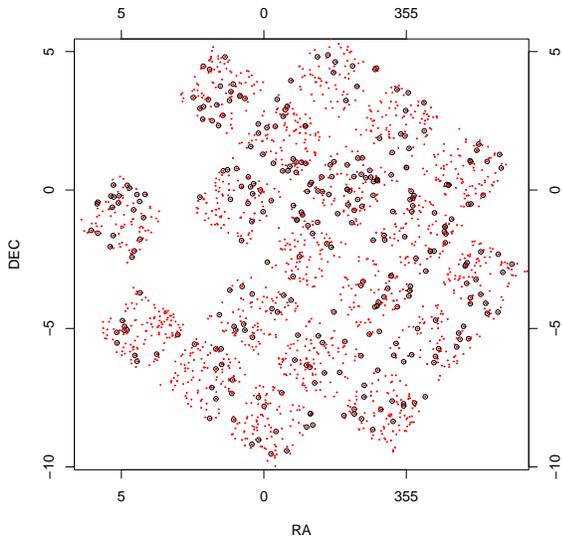}
\caption{Distribution of the 2079 K2-E2 targets (small red dots) and our subsample (small black circles) in the K2-E2 field.}
\label{fig:radec}
\end{figure}

The ecliptic latitude of the stars spanned $\beta$= -8.6$\degrees$ --  +5.5$\degrees$, which is representative of the K2-E2 and future K2 targets. Fig.~\ref{fig:radec} shows the uniform distribution of our subsample in the K2-E2 field.

\section{Effect of asteroids on K2 photometry}\label{effects}

\begin{figure}
\centering\includegraphics[width=6.2cm,angle=270]{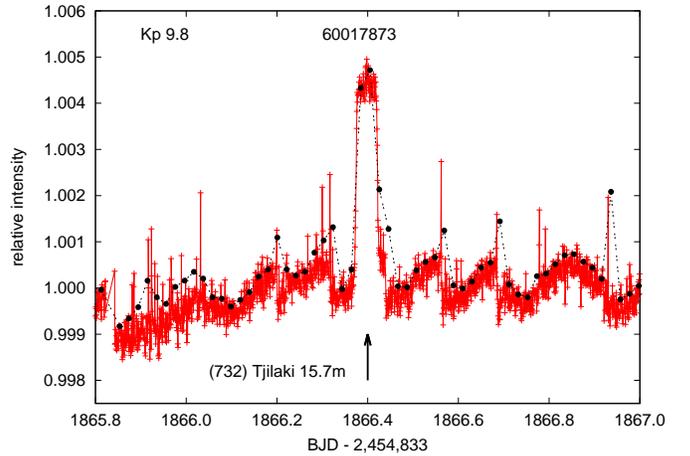}
\caption{Light curve of the encounter with (732)~Tjilaki. The magnitude of the star is in the upper left corner. The ID of the target star is in the middle. Red dense points are short cadence observations, black points are long cadence data points. A few outlier (bright) data points were removed from the short cadence light curve. The asteroid crossing event is marked by an arrow. The ID and the brightness of the asteroid can be found next to the arrow.}
\label{fig:lc1}
\end{figure}

\begin{figure*}
\centering\includegraphics[width=15.5cm,angle=0]{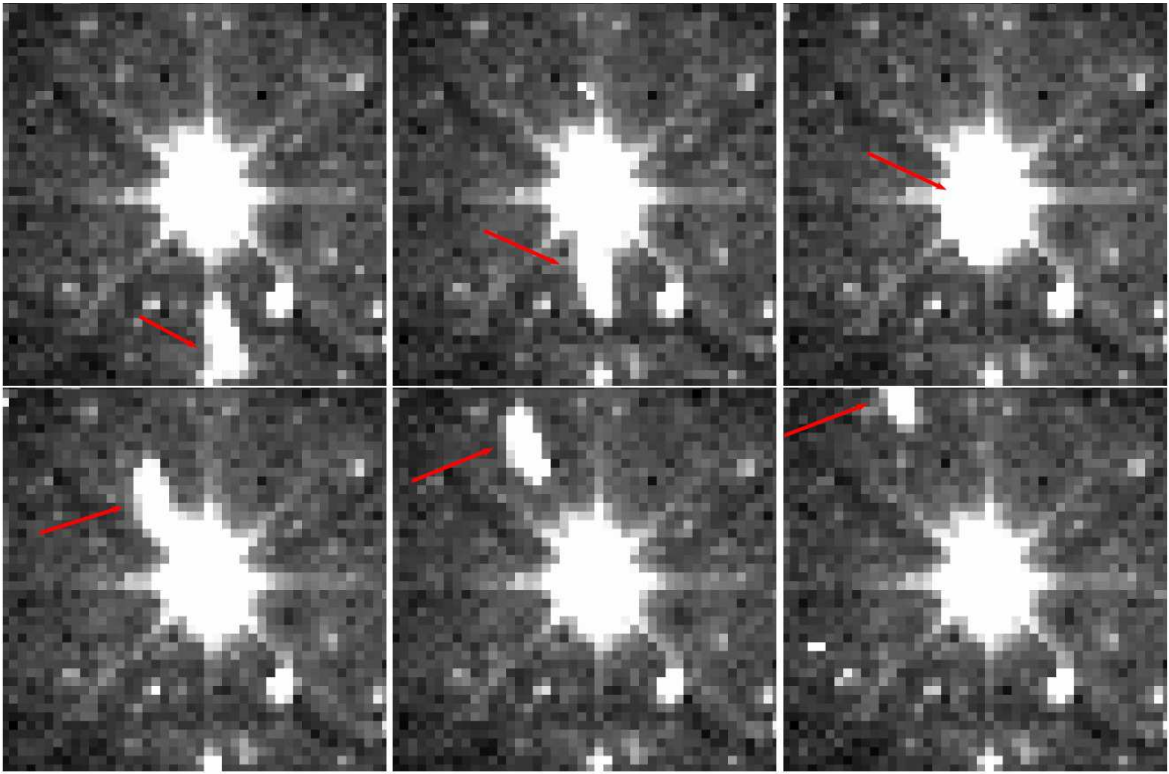}
\caption{Close encounter of the 15.7 magnitude asteroid (732)~Tjilaki with the K2 target 60017873 (Kp=9.8 mag) in the K2 engineering test observations, close to the ecliptic. The frames are taken in long cadence mode and span three hours. The asteroid - shown by a red arrow - starts as a vertically elongated object at the bottom of the image, and left the pixel mask on the top of the image. Time increases from left to right, from the top to the bottom.  See Fig.~\ref{fig:lc1} for the light curve of the same event.}
\label{fig:encounter}
\end{figure*}

In Fig~\ref{fig:lc1} we show the light curve of the encounter of the relatively bright asteroid 732 (Tjilaki), with the star EPIC 60017873, which is a bright target (Kp=9.8). The images of the same event are presented in Fig~\ref{fig:encounter}. We choose to follow a typical aperture photometric procedure foreseen for the K2 field observations. We applied a 14x14 pixel mask for simple aperture photometry. The local background was removed.  We also removed 64 outlier points from the short cadence time series, which corresponds to less than 0.5\% of the total number of exposures points (13088). These points are egregiously bright outliers, standing out by more than $10\thinspace \sigma$ from the light curves. These anomalies most probably correspond to the  pointing adjustments performed every six hours, based on their timing and their common appearance in all light curves we investigated. We did not apply any decorrelation with external parameters, nor trend filtering, thus our figures show the original, 'raw' K2 light curves, except a few outliers that were removed. We chose this star because it was observed in short cadence mode as well (red points). The additional flux from the asteroid is clearly seen. We note that the long cadence data points also show the flux excess, and the duration of the event makes it possible to differentiate between cosmic ray or other short time-scale instrumental events and asteroid encounters. We did not remove all the LC points that contain a removed SC photometric data point in order to prevent the removal of too many LC data. This led to a slight discrepancy between the LC and the SC light curve which can be noticed in Fig~\ref{fig:lc1}, but has no consequences for the results that follow.

\begin{figure*}
\centering\includegraphics[width=4.cm,angle=270]{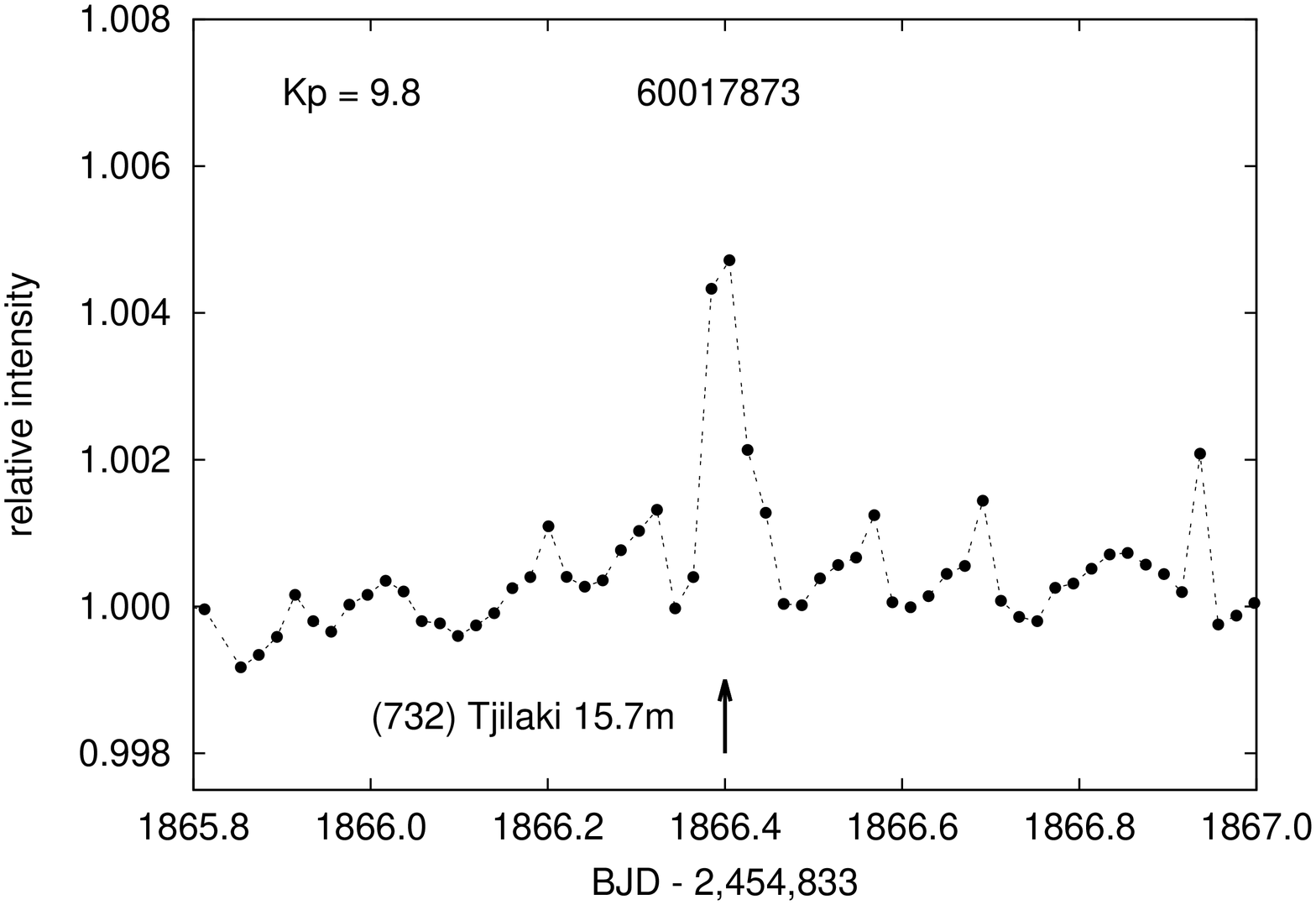}
\centering\includegraphics[width=4.cm,angle=270]{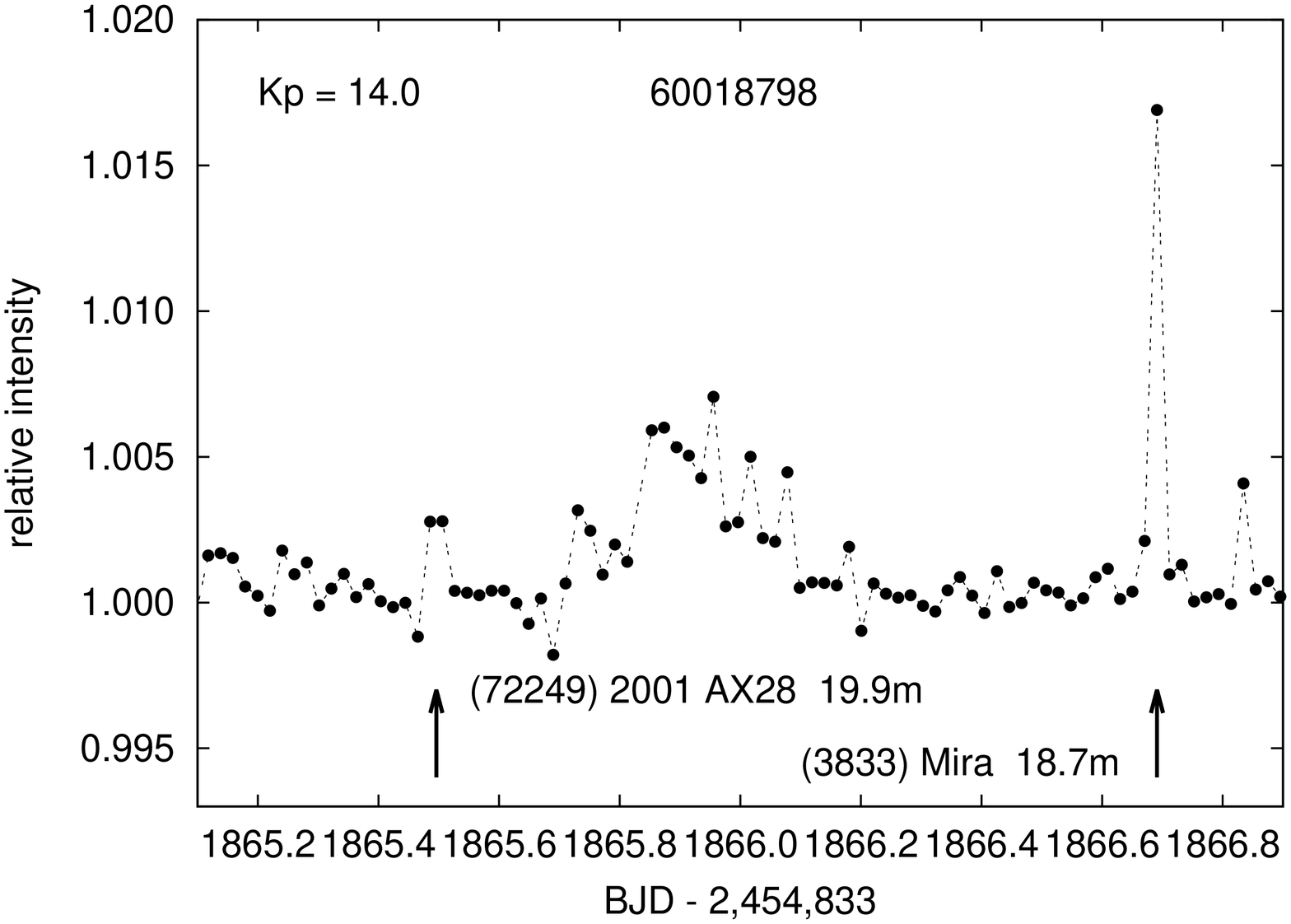}
\centering\includegraphics[width=4.cm,angle=270]{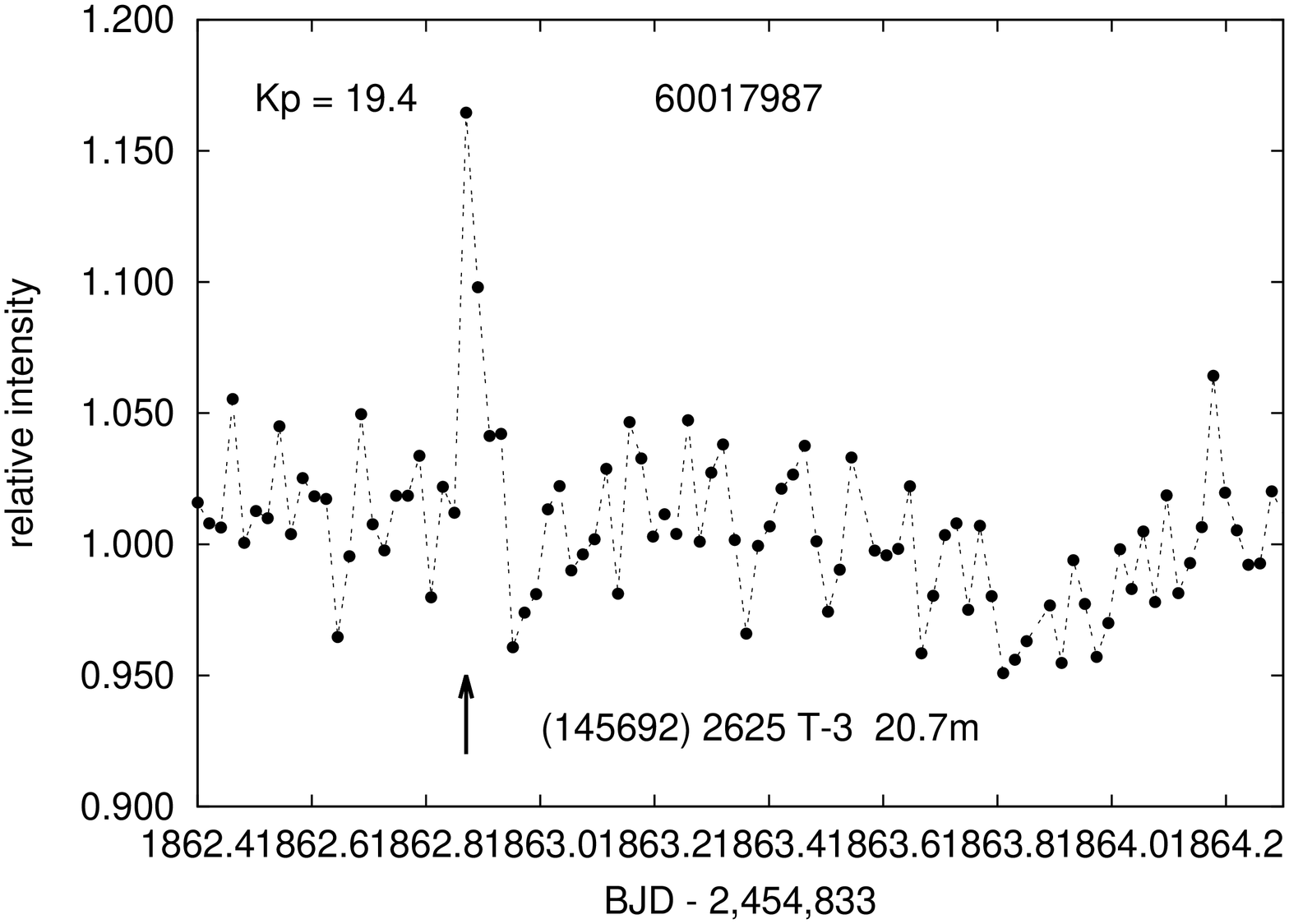}
\caption{Long cadence light curves of encounters of asteroids and K2 targets. The notations are the same as in Fig~\ref{fig:lc1}. From left to right we plot a bright, an average brightness and a faint target star.}
\label{fig:lc2}
\end{figure*}

In Fig~\ref{fig:lc2} we show three examples with targets of different brightness (Kp=9.8, 14.0 and 19.4 from left to right), all showing one or more asteroid encounters. It is easy to see that very faint asteroids can cause outlying photometric points when our target is faint.

\section{Identification of the observed asteroids}\label{ident}

In order to efficiently identify the apparent ``flow'' of minor planets in the fields, we computed their standard J2000 heliocentric ecliptic celestial positions  based on the orbital data provided by MPC\footnote{http://www.minorplanetcenter.net/iau/MPCORB.html} directly. As {\it Kepler} has a heliocentric, Earth-trailing orbit, we need to compute the directions as seen from {the spacecraft. The actual coordinates of the {\it Kepler} spacecraft have been retrieved from Horizons\footnote{http://ssd.jpl.nasa.gov/horizons.cgi}. The retrieval has been performed on a fine sampling (1 days of step size) while intermediate coordinates have been computed using cubic spline interpolation. Then we used the {\it Kepler} coordinates as the origin for the computations.}

The heliocentric positions of the minor bodies have been computed using numerical integration based on the initial conditions provided by the MPC database. The integration scheme is based on Lie-series \citep{pal2007} which provides both a fast method and an easy way of precise interpolation between subsequent steps, i.e. lacking complex series of operations. During integration, restricted N-body dynamics have been considered.

Apparent positions are then computed including light-time effect and proper handling of TT - TAI - UTC differences. Compared with outputs of the Horizons service, the aforementioned procedure yields the same series of coordinates within 0.3" (i.e. 0.0001 degrees). Hence, safe cross-identification of known moving objects is quite effective and reliable down to within an arcsecond. This accuracy does not practically yield any ambiguity even at low ecliptic latitudes, where the apparent density of such objects is the largest. Even if we neglect mutual perturbations in the computations, the uncertainty is still 2-3 pixels in the Kepler FOV, ensuring unambiguous identification. In addition, the procedure does not require to flood external services: it is a prominent issue when one has to handle several hundreds of disjoint microfields in the same run.

Among the 232 asteroids that we found, all of them were known and numbered objects except one, which is 2013~OE, a Near-Earth Object. 2013~OE was identified only on two frames due to its high velocity. Based on this result and our previous experiences to find asteroids in this magnitude range, we conclude that only a few new main-belt asteroid discoveries are expected during the K2 Mission, the majority of the observable asteroids will be already known objects. 

\section{Asteroid photometry with K2}\label{phot}

Here we demonstrate that the K2 mission is capable to measure the light curves of asteroids in microfields above 50--100 pixel diameter, and of course in observations of larger swaths of pixels, possibly leading to several thousand asteroid light curves in the entire mission. To date, rotation periods of $>$10,000 asteroids are known \citep{szabo2008}, but very few asteroids are found below the limiting magnitude of 17-18, and only 170 of them have reliable spin axis determinations (Asteroid Photometry Catalog\footnote{ http://asteroid.astro.helsinki.fi}, \citealt{laspina2004}, \citealt{kryszczynska2013}). Thus, rotation periods for asteroids fainter than H$\approx$13--14~mag are mostly known for a spatially (and rotationally, \citealt{laspina2004}) selected sample of NEOs. For a significant number of asteroids, however, where photometric data have low coverage, ground based observations do not provide a unique period, but a set of alias periods instead.

\begin{figure}
\centering\includegraphics[width=7.cm,angle=270]{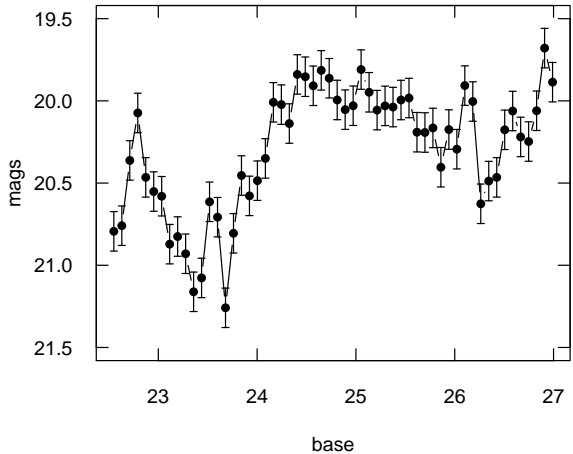}
\caption{The light variation of (120934) 1998 SE149 during a 4.5 hour observing run.}
\label{fig:lc_ast}
\end{figure}

In this task, the main power of K2 mission is the continuous observation of main-belt asteroids in the fields, leading to $\sim$40~h continuous coverage for full frames. The limiting magnitude in K2 asteroid photometry is $\sim$2~mag below the current databases of asteroid light curves that are sufficient for reliable period determination \footnote{ E.g. see the Asteroid Photometry Catalog and the Asteroid Lightcurve Database available at  http://www.minorplanet.info/lightcurvedatabase.html}.

In this exercise we used short-cadence data. In Fig.~\ref{fig:lc_ast} we present the light curve of asteroid (120934) 1998~SE149 that exhibited 20.4~magnitude  brightness at 2.917~AU heliocentric radius and at 3.222~AU distance from the observer, i.e the Kepler spacecraft. The object passed through the 50$\times$50 pixel area on 3/4 February, between 22.55--02.99~UT. To increase S/N, five consecutive images were co-added. Since the asteroid trailed during the time of five exposures, photometry was performed in a 3$\times$3 aperture, centered on  the brightest pixel of the asteroid image.

The most sensitive task is the subtraction of the background, especially for an object near the faint limit. The difficulty is that the pointing of the Kepler spacecraft varies a little, therefore the image position of stars vary in the sub-pixel range. This is enough to disable the construction of a global template image of the background that can simply be subtracted from the observed images. Instead, we determined the similarity between the processed image and all other images by a cross-correlation, masking out the brightest (saturated) star and the asteroid itself. Then we selected the most similar 30\% of images, and calculated their median. This way, the stars could have been satisfactorily removed whenever there were enough similarly pointed observations in the data set.

The asteroid crossed the image of the star 60017839 (Kp=13.9). The light curve shows two humps (probably when the asteroid is seen from the sides) and a fainter state in between. There is no clear periodicity, while the general trend in the light curve suggests that the rotation period exceeds the 4.5~h of observation. The scatter of the light curve was estimated in the shallow local minima (between the 23th and 40th stacked image). After removing the linear trend, the standard deviation of photometric points is 0.12~mag during 5 minutes of observation. 

This observation therefore demonstrates that an acceptable quality photometry can be performed in K2 images down to around 20.5~magnitude. Since we expect 2 days of observability of main-belt asteroids in K2 full frames, and the same time range for slowly moving trans-Neptunian objects (TNOs) in few-100-pixel-size apertures, the $a/b$ shape asphericity can be fitted with a better accuracy than 0.5\% and the period can be securely detected if the brightness variation of the asteroid exceeds about 0.01~mag, down to 20.5 limiting magnitude. The estimate was based on a $>$2-days observation at the 0.12 mag scatter level, accounting for 96 photometric points. Assuming a white noise, the expectation value of the noise then will be 0.01, therefore a 0.01 mag amplitude variation will emerge as an S/N=1 peak. When we can expect 5 days of observability of main-belt asteroid, their rotation period can be detected at an S/N$>$2 level down to 20.0 limiting magnitude, if the brightness variation of the asteroid exceeds about 0.01 magnitude.

\section{Summary}\label{sum}

This work demonstrates for the first time that apparently close encounters of main-belt asteroids with photometric targets will be very common in the Ecliptic K2 Mission. We examined 300 of the 2079 publicly available light curves and found that in around half of them the effect of asteroids cause brightness increase. The typical length of this event (when brightening occurs due to the large aperture that is usually applied to extract photometry) is less than three hours. The typical brightness of the asteroids we found are 18-21.5 (the faintest is 21.7) magnitude. We successfully identified all of our Solar System bodies, because this engineering field was pointed around 70 deg elongation, where the asteroids are 1.6-3.4 mag (in the r=3.5-1.9 AU heliocentric distance range) fainter than they opposition magnitude, and our asteroid sample is complete down to about opposition magnitude 20. In principal it would be possible to discover new asteroids with the K2 Mission, because the observations are typically taken starting from an elongation of 135 deg, where the main-belt asteroids are fainter than their opposition brightness only by 0.8-1.3 magnitude.
However, as the measurements are taken on the evening sky (as seen from Earth), after the opposition point, and as the 1.5-1.8 m telescopes of the ongoing large asteroid searching surveys, like the  Mt. Lemmon and the Pan-STARRS surveys search the vicinity of the opposition point down to 21.5 magnitude, we do not expect a  significant number of new main belt asteroid discoveries from K2.
We plan to publish the details of our identification process in a forthcoming paper. Lastly, we demonstrated that accurate photometry of asteroids is possible. The process will be particularly relevant in large contingent fields of the K2 ecliptic fields, where thousands of main-belt asteroids are expected, and in future space photometric missions, such as  TESS \citep{ricker2014} and PLATO \citep{rauer2014}, where ecliptic fields will be also targeted.  

\begin{acknowledgements}
{This project has been supported by the `Lend\"ulet-2009 and LP2012-31 Young Researchers' Program of the Hungarian Academy of Sciences, the Hungarian OTKA grants  K-83790, K-109276 and K-104607 and by City of Szombathely under agreement no. S-11-1027. The research leading to these results has received funding from the European Community's Seventh Framework Programme (FP7/2007-2013) under grant agreements no. 269194 (IRSES/ASK), no. 312844 (SPACEINN), and the ESA PECS Contract No. 4000110889/14/NL/NDe. Gy.M.Sz. was supported by the J\'anos Bolyai Research Scholarship of the Hungarian Academy of Sciences. We thank L. Moln\'ar for fruitful conversations on K2 photometry and the referee for insightful comments that improved the quality of the paper.}
\end{acknowledgements}


\end{document}